\documentclass[preprint,12pt]{elsarticle}
\usepackage{graphicx}
\usepackage{subfigure}
\usepackage{epsfig}
\usepackage{color}
\usepackage{comment}
\usepackage{enumerate}
\usepackage{amsmath}
\usepackage[T1]{fontenc}
\usepackage{amssymb}

\journal{Physica A}

\begin{document}

\begin{frontmatter}



\title{Delays-induced Phase Transitions in Active Matter}


\author[inst1]{Fatemeh Pakpour}

\affiliation[inst1]{organization={Science Department},
            addressline={Arak University of Technology}, 
            city={Arak},
            postcode={3818146763}, 
            state={Markazi},
            country={Iran}}

\author[inst2,inst3]{Tam\'{a}s Vicsek}

\affiliation[inst2]{organization={Department of Biological Physics, E\"{o}tv\"{o}s University},
            addressline={ P\'{a}zm\'{a}ny P\'{e}ter s\'{e}t\'{a}ny 1A}, 
            city={Budapest},
            postcode={H-1117},
            country={Hungary}}

\affiliation[inst3]{organization={MTA-ELTE Statistical and Biological Physics Research Group of HAS},
            addressline={ P\'{a}zm\'{a}ny P\'{e}ter s\'{e}t\'{a}ny 1A}, 
            city={Budapest},
            postcode={H-1117},
            country={Hungary}}            

\begin{abstract}

We consider the patterns of collective motion emerging when many aligning, self-propelling units move in two dimensions while interacting through a  repulsive potential and are also subject to delays and random perturbations. In this approach, delay plays the role analogous to reaction time so that a given particle is influenced by the information about the velocity and the position of the other particles in its vicinity with some time delay. To get insight into the involved complex flows and the transitions between them we use a simple model allowing – by fine-tuning of its few parameters – the observation and analysis of behaviours that are less accessible by experiments or analytic calculations and at the same time make the reproduction of experimental results possible. We report for the first time about a transition from an ordered, polarized collective motion to disorder as a function of the increasing time delay. For a fixed intermediate value of the delay similar transition (from order to disorder) is obtained as the repulsion radius is increased. Our simulations show a transition from total polarization to two kinds of states: fully disordered and a kind of state which is a mixture of patches of fully disordered motion in the background of orderly moving other particles. The transition occurs as the delay time is increased and is sharp, indicating that the nature of this order-disorder transition is either of first-order or is described by a sharply decreasing linear function. Our model is a simplified version of a practical situation of quickly growing interest because time delays are expected to play an increasingly important role when the traffic of many, densely distributed autonomous drones will move around in a quasi-two-dimensional air space.

\end{abstract}


\begin{highlights}
\item Almost all systems of collectively moving units are subject to some sort of delays caused by finite time intervals needed for reacting to the positions and velocities of the others. Still, a systematic study of the effect of delays on the destruction of coherent flocking is lacking. Here we show that if the members of a group of aligning particles travelling in two dimensions have a finite size, i.e., repel each other, a large enough reaction time results in either a complete disorder or in a new kind of motion state in which patches of disordered and ordered states can coexist. Our findings are expected to be relevant to situations such as the traffic of many units, e.g., drones in a confined area.
\end{highlights}


\end{frontmatter}


\section{Introduction}
\label{sec:sample1}

Systems of many similar units exhibit collective behavioural patterns through interactions among their units. In the case of collective motion, these interactions can be looked at as a result of exchanging information about the positions of the others and then accompanied by the corresponding reaction. In nearly all systems the reaction to a particular set of information about the positions of the others is not instantaneous due to several sources of non-negligible reaction times resulting in delays. The mechanisms behind delayed reactions are common in both living and technological systems. In this work, we explore in which ways time delays can result in the destruction of coherent behaviour during the motion of aligning self-propelled particles (SPP-s) representing a typical manifestation of active matter.

The study of collective motion \cite{Vicsek2012} of groups of biological and non-biological agents involved also in the context of soft active matter \cite{Marchetti2013} has been a subject of quickly growing interest over the past two decades. Examples of related studies include
bacterial colonies \cite{ Zhang2010}, cell migration \cite{Mehes2014}, swarms of insects\cite{Buhl2006}, schools of fish \cite{Lopez2012}, flocks of birds \cite{Bajec2009}, and crowds of people \cite{Moussaid2010} to nematic fluids \cite{Doostmohammadi2018}, shaken metallic rods \cite{Kudrolli2010}, and flying robots \cite{Viragh2014}. A comprehensive understanding of the complex and nonlinear dynamics of a system consisting of a large number of interacting entities in addition to environmental impacts and important applications in biology \cite{Deutsch2012}, ecology \cite{Westley2018}, ethology \cite{Lopez2012, Sumpter2010}, agriculture \cite{Buhl2006}, economics \cite{Cont2000}, social science \cite{Helbing2000}, and cognitive science \cite{Couzin2009} can be functional in developing modern technologies of artificial agents such as drones \cite{Vasarhelyi2018} and autonomous flying robots \cite{Brambilla2013, Turgut2008}. The fascinating and extraordinary features of the collective motion in diverse systems of interacting units over a wide range of system sizes such as self-organization without any centralized control and leaderless coherent motion \cite{Aoki1982, Gregoire2003}, hierarchical leader-follower relationships among the members \cite{ Nagy2010, Couzin2005}, long-lasting giant number fluctuations \cite{Chate2006, Narayan2007}, turbulent flow and vortex formation \cite{Cisneros2007, Sokolov2009, Wolgemuth2008}, and transitions to a variety of motions namely marching, milling, rotating chains, and jamming \cite{Kudrolli2008, Helbing2000, Ihle2011, Mishra2010, Tarcai2011, Helbing2009} have been manifested through numerical and experimental investigations which can be interpreted in the scope of equilibrium and non-equilibrium statistical physics. 

Various mathematical and computational models \cite{Aoki1982, Reynolds1987, Vicsek1995, Toner1995, Shimoyama1996, Couzin2002, Toner2005, Cucker2007, Derzsi2009, Cavagna2018} have been introduced to investigate the dynamics of self-propelled particles. The simplest model accounting for the non-equilibrium phase transition in flocking as a function of random perturbations was introduced by Vicsek et.al. \cite{Vicsek1995} which we shall refer to as SPP95. In the SPP95, the collision avoidance effect is ignored and aligning with neighbouring particles occurs instantly. However, in realistic systems, either natural or artificial, in addition to the requirement of considering repulsive interactions to prevent collisions, each agent needs some time to send, receive and respond to the information signals from its neighbourhood. Thus, although the very simple SPP95 model already displayed a non-trivial non-equilibrium phase transition as a function of the level of random perturbations it was not designed to account for the above-mentioned details of the interactions potentially resulting in destroying the polarized motion which usually emerges if only alignment is considered.

One of the very early studies considering the problem of delay time in the collective behaviour of many-body systems appeared in the subject of vehicular traffic. It was known that the observed density waves and instabilities in vehicle traffic are not explained without considering the reaction time of drivers \cite{Chandler1958, Kometani1958, Herman1959}. Following that, the question of sensory temporal delay provided a solution to the problem of gradient sensing mechanisms in bacterial colonies. An experimental research showed that bacterial chemotaxis appears using a time-sensing mechanism as if they have a kind of "memory device" to compare the difference between environmental concentrations at successive times \cite{Macnab1972}. The effect of time delay was also investigated in different many-body physical systems such as physiological control systems \cite{Mackey1977}, the ring cavity system \cite{Ikeda1979}, cortical neurons \cite{Landsman2007}, and lasers \cite{Carr2006} which showed oscillatory behaviours and chaos.

The emergence of instability in a two-dimensional system of self-propelled agents interacting via a pair-wise linear attractive force and in the presence of noise and a constant time delay was demonstrated in \cite{Forgoston2008}. The study showed that the delay induces a transition into a state which is an oscillating aligned swarm that depends on the size of the interaction coupling coefficient. In a work reporting on both a simulation of flocking birds with two decentralized control algorithms and an experiment with real autonomous flying robots, the realistic features of the flocking such as time delay, inertial effects, sensor errors, and locality of communication besides a friction-like interaction for alignment of the neighbouring particles, a repulsive force, and noise considered \cite{Viragh2014}. The authors showed that although time delay can reduce the stability of the flocking state the friction-like term reduces the strength of this instability. They also demonstrated how time delay leads to oscillations in the system and indicated that an additive Gaussian noise term can decrease the instabilities caused by delay time.

In a one-dimensional numerical study on the directional switching of the self-propelled agents with delayed interaction, it was shown that with increasing delay time the average switching time of the particles increases \cite{Sun2014}. The authors examined three types of delay time: a particular time delay for all particles, a random delay chosen from a normal distribution, and a state-dependent form of delay time in which it increases with the level of disorder in the vicinity of the particle. They showed the mean switching time for the group with identical delay time is shorter than those with random delay time and longer than the third group with heterogeneity. They also argued that the time delay may facilitate the coherent motion of the latter group. 

The effect of delay on the collective behaviour was also investigated experimentally by using a small number of autonomous phototactic robots being able to measure the intensity of a decaying light field emitted from the others \cite{Mijalkov2016}. It was shown that sensorial delay - the time each particle needs to change its velocity in response to the intensity of light received from another particle - can be a control parameter for the type of collective movement of particles from segregation to aggregation and clustering. To confirm the scalability of the used mechanism, they simulated the behaviour of a group of 100 robots and showed a positive delay results in forming clusters while a negative delay can cause the robots to move away from each other. Another numerical study on the role of time delay on the safety of an assembly of up to 5 Unmanned Aerial Vehicle (UAVs) in a 3D space was conducted \cite{Casas2018}. The research showed that communication delay has a major role during the flock-building stage; longer delay time resulted in an extended time needed for the flock to stabilize in the desired direction. The authors also showed a very high delay time leads to an oscillatory behaviour which in turn increases the possibility of collisions and decreases the safety of the group. They argued that the number of agents building a flock is limited by the duration of the communication delay. 

The facilitation of coherent motion for a group of self-propelling agents at short delay times was also demonstrated in a two-dimensional numerical investigation based on the SPP95 and using a Voronoi tessellation tool for more efficient measurement of clustering formation \cite{Piwowarczyk2019}. In contrast, the study showed the order is destroyed when the delay time is around 25 times longer. This phenomenon was also observed in another numerical work using the SPP95-like model of active matter in three dimensions with periodic boundary conditions \cite{Holubec2021}. The authors showed the delay time first facilitates the ordered motion but with increasing time delay the characteristic oscillation of delayed systems is observed. They observed an identical finite-size scaling and correlation as natural swarms show. They also argued that the effect of time delay in a system of SPPs is similar to inertia.

Although in practical situations the magnitude of the velocities of the SPPs can vary over a broad range, in most of the related simulations the speed of the particles was assumed to be a constant. A particularly detailed/complete model  \cite{Hemelrijk2015} took into account numerous realistic factors playing an essential role in the formation of flocks of birds. Their StarDisplay model and the related simulations take into account the reaction times as well and are capable of reproducing many realistic flocking scenarios. Perhaps the most relevant example, in which the speed of the flocking units has to be considered is a flock of drones moving in confined areas. Such a system was investigated both numerically and experimentally in a study in which over 30 autonomous drones were used to explore the effects emerging due to the criterion of avoidance (drones were not allowed to move too close to each other) and the time delays resulting from the onboard calculations carried out by the robots \cite{Vasarhelyi2018}. It was demonstrated that only a careful (evolutionary) optimization of the parameters of the autopilot code could result in a coherent and collisionless motion under such conditions.

In the present work, we focus on the statistical mechanics effects of taking into account the two realistic conditions of delays and repulsion. Instead of looking for reproducing observed patterns of flocking, we consider the general question of a possible phase transition in the systems of SPPs as a result of increasing delay times and repulsion radii.

\section{Model, Parameters, and Definitions}

In order to investigate the effect of time delays in systems in which they may play a significant role, we consider $N$ particles moving continuously in an $L \times L$ cell with periodic boundary conditions. At $t = 0$, the particles are randomly distributed within an area of a given size at positions $\mathbf{x}_i(0)$. In addition, each particle is assigned a velocity $\mathbf{v}_i(0)$ with the same magnitude,  $v_o$, and randomly distributed directions $\theta_i(0)$. The time interval between two updates of the positions and directions is $\Delta t = 1$. At each time step, the position of the $i^{th}$ particle is updated according to:

\begin{figure}[htbp]
\includegraphics[width=5.5in]{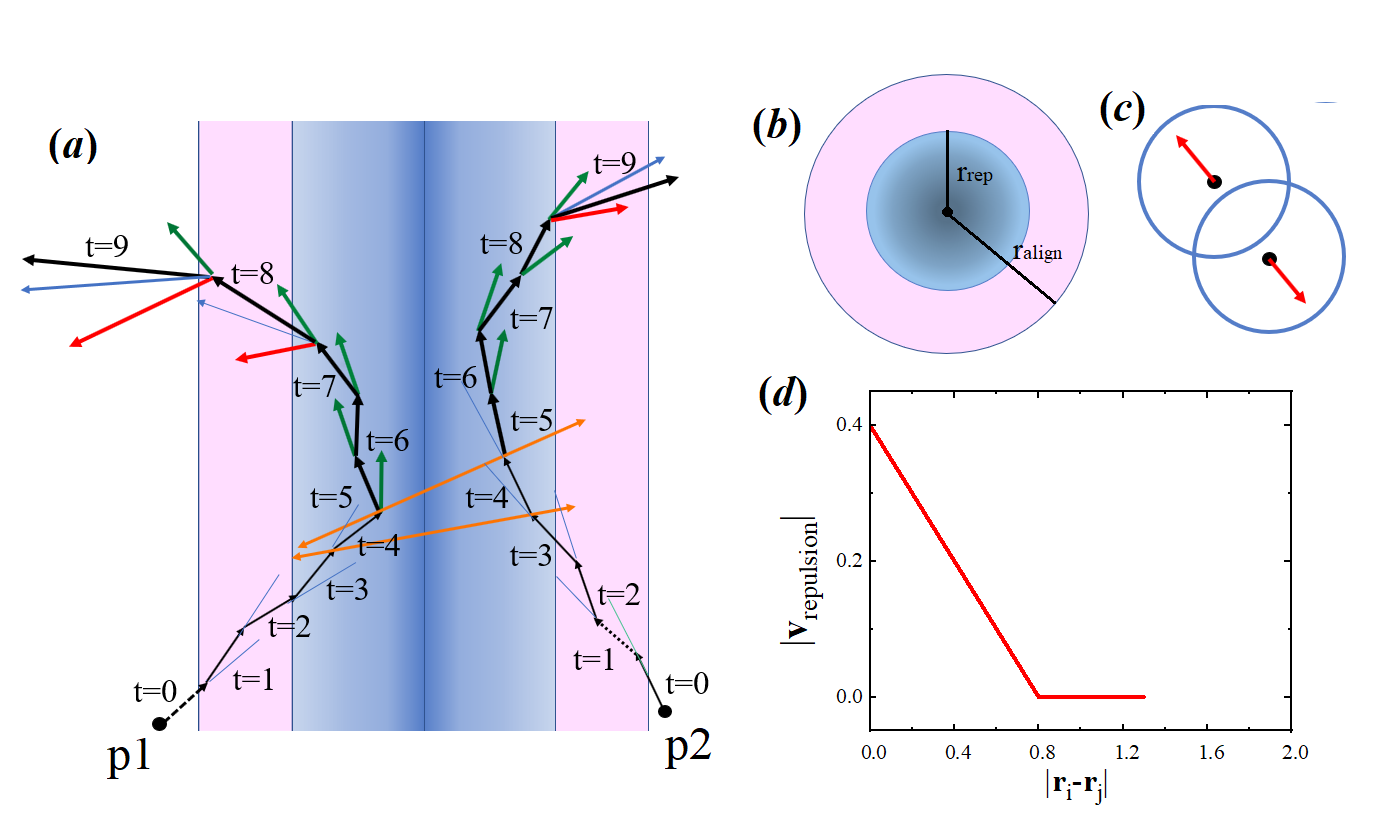}
\caption{Visualization of the model. $\mathit{(a)}$ is an explanatory panel, demonstrating the way in which delay plus repulsion can result in a larger angle between the departing particles as compared to the angle between their trajectories before "collision". The interaction starts at $t=4$ and the delay time for particle1 $\tau_1=3$, and for particle2 $\tau_2=4$. At $t=4$ particle1 remembers the information from particle2 at $t=1$ (the dotted velocity vector), while particle2 remembers the information from particle1 at $t=0$ (the dashed velocity vector). For each step, the green vector is the alignment term, $\mathbf{v}^{align}$, the red one is the repulsion term, $\mathbf{v}^{rep}$, the blue vector is the sum of two terms, and the black vector is the final velocity of the particle after adding random noise to the blue one. The repulsion term is in the direction of the distance vector between two particles that was shown with the orange vector. $\mathit{(b)}$ shows the repulsion and alignment radii, $\mathit{(c)}$ illustrates the repulsion force direction, $\mathit{(d)}$ displays the mathematical form of the repulsion force. Supplementary movie 1 complements the above visualization by illustrating the behaviour of the particles on a small scale compatible with the range of the alignment radius.}
\label{Fig1}
\end{figure}

\begin{equation}
\mathbf{x}_i(t+1)= \mathbf{x}_i(t) + \mathbf{v}_i(t+1) \Delta t, 
\end{equation}

\noindent
where $\mathbf{x}_i(t)$ is the position at time $t$ and $\mathbf{x}_i(t+1)$ and $\mathbf{v}_i(t+1)$ denote the position and the velocity of the agent at time $t+1$. We use the term agent and particles interchangeably and a collection of aligning agents represents active matter. The velocity of each agent at time $t+1$ is calculated by considering three terms: i) the effect of aligning with neighbouring particles located within a certain alignment radius, $R$, ii) the repulsion from particles closer than a certain repulsion radius, $2r_{rep}$, and a random noise $\eta_i(t)$. Each particle has a delay time, $\tau_i$, it spends on processing information about the positions and the velocities of the others, while it reacts without delay regarding information about itself. The delay times in this simulation were chosen from a Gaussian distribution \cite{Islam2023} with mean $\mu_{\tau}$ and standard deviation $\sigma_{\tau}$.

The alignment term for $i^{th}$ particle at time $t+1$ is calculated by taking the average over the particle's velocity at time $t$ and the velocity of neighbouring particles located within radius $R$ at time $t-\tau_i$:

\begin{equation}
\mathbf{v}^{align}_i(t+1) = \mathcal{N}^{v_o}\left(\frac{\sum_{j=1}^N a_{ij}(t-\tau_i)\mathbf{v}_j(t-\tau_i) + c_{align} \mathbf{v}_i(t)}{s_i(t-\tau_i) + 1} \right),
\end{equation}
\noindent
where $\mathcal{N}^{v_o}()$ is a function which normalizes the magnitude of its argument to $v_o$ and $s_i(t-\tau_i)$ is the number of neighbours inside radius $R$ at time $t-\tau_i$. $c_{align}$ is the coefficient of alignment and $a_{ij}$ are the elements of the alignment matrix defined as follows:

\noindent
\begin{flalign}
a_{ij}=\begin{cases}
          1, \ \  i\neq j, \ \ i=1,...,N,\ \ j\in S_i(t) \\
          0, \  \  \text{otherwise,}
       \end{cases}   
\end{flalign}
 \\
\noindent 
where $S_i(t)$ is the set of particles inside the interaction radius $R$ and for simplicity we assumed the simplest form of the contribution of the velocity of a particle to the resulting direction (all $a_{ij}=1$ or $0$). The alignment matrix is updated at each time step.  

The repulsion term is a sum of pairwise repulsion forces depending on the distance of particles according to: 

\begin{equation}
\begin{split}
\mathbf{v}^{rep}_i(t+1) = c_{rep} \sum_{k}  g(r_{i} + r_{k} - \lvert\mathbf{x}_{ik}(t-\tau_i)\rvert)\frac{\mathbf{x}_{ik}(t-\tau_i)}{\lvert\mathbf{x}_{ik}(t-\tau_i)\rvert},
\end{split}
\end{equation}
where $c_{rep}$ is the repulsion coefficient, $r_{i}$ is the repulsion radius of particle $i$ and $r_k$ is the repulsion radius of the neighbouring particle $k$ and $k$ runs over the particles located within a circle centred at particle $i$ with the radius of $r_i + r_k$. We assume that both the $r_{i}$ and $r_{k}$ are equal to $r_{rep}$. $\mathbf{x}_{ik}(t-\tau)$ is $\mathbf{x}_i- \mathbf{x}_k$ at time $t-\tau$. The form of $g(x)$ can be chosen quite freely, however, for the sake of simplicity we shall assume a simple hard-core type dependence, meaning that repulsion acts only within a given repulsion radius. Thus we choose the g(x) function as follows: 
\noindent
\begin{flalign}
g(x)=\begin{cases}
          0, \ \  \  |\mathbf{x}_i - \mathbf{x}_k| \geq r_i + r_k \\
          x, \  \  \  \text{otherwise.}
       \end{cases}   
\end{flalign}

\noindent
Finally, the velocity vector at time $(t+1)$ is obtained as follows:

\begin{equation}
\mathbf{v}_i(t+1) = \mathcal{R}^{\eta_i(t+1)}\left[ \mathcal{N}^{v_{max}}\left(\mathbf{v}^{align}_i(t+1) + \mathbf{v}^{rep}_i(t+1) \right)\right],
\end{equation}

\noindent
where again $\mathcal{N}^{v_{max}}()$ normalizes its argument to a maximum magnitude, $v_{max}$, and the function $\mathcal{R}^{\eta_i}()$ is a rotation operator that changes the angle of its argument with $x$ axis by the amount $\eta_i$ where $\eta_i$ is a random number chosen from a uniform distribution from $[-\eta/2,\eta/2]$ interval. A visualization of the model is shown in Figure \ref{Fig1} and a list of parameters used in our model is given in Table \ref{table:1}.

\begin{table}[h!]
\centering
\caption{The list of parameters together with their explored ranges used in the simulations of the model of flocking with repulsion and delay.} 
\vspace*{5mm}
\begin{tabular}{ |p{2cm}|p{8cm}|p{2.5cm}|  }
 \hline
 \multicolumn{3}{|c|}{List of Parameters} \\
 \hline
 Parameter&Definition&Range\\
 \hline
 $N$&    Number of particles   &$2000-16000$\\
 $\rho$    &Number density of the particles&   $1.0 - 3.0$\\
 $v_o$&    Initial velocity    &$0.01-0.03$\\
 $R$&    Alignment radius   &$1.0-1.5$\\
 $c_{align}$    & Alignment coefficient&  $1.0-3.0$\\
 $r_{rep}$  & Repulsion radius&  $0.2-0.45$\\
 $c_{rep}$&   Repulsion coefficient&$0.1-1.0$\\
  $v_{max}$&   Maximum value of the velocity &$3v_o$\\
 $\eta$ & Maximum noise angle& $0 - \pi$\\
 $\mu_{\tau}$& Mean of time delay distribution  &$0 -40$\\
 $\sigma_{\tau}$&  Standard deviation of the delay distribution& $ \mu_{\tau}/5, \mu_{\tau}/10$\\
 $T$&  The number of iterations in any run& $6000-14000$\\
 \hline
\end{tabular}
\label{table:1}
\end{table}

To describe the degree of global alignment in the system the normalized order parameter is calculated using the following equation at each time step \cite{Vicsek2012}:

\begin{equation}
\noindent
\Phi(t) = \frac{1}{N} \bigg|\sum_{j=1}^N \frac{\mathbf{v}_j(t)}{\lvert\mathbf{v}_j(t)\rvert}\bigg|.  \\
\end{equation} 

\noindent
The value of $\Phi$ varies from nearly zero in a situation of disordered motion and tends to $1.0$ for ordered/polarized motion.

\section{Simulation and Results}

The simulations were carried out using a GPU-accelerated code based on CUDA Python making use of parallel processing. Each point in all plots was obtained from averaging over 5 separate runs. The dynamics and the degree of coordination in the system were tracked using equations $1-7$ through several thousands of time steps depending on the system size. To account for the effect of time delay we first allowed the particles to move freely from the initial state for a short interval, normally a few time steps longer than the maximum time delay, and then the new positions and velocities were calculated. As mentioned in the previous section the time delays were drawn from a normal distribution so that they were different for different particles. We explored a wide range of parameters to find the most relevant patterns of the motion of particles. The behaviour of the system and the phase transitions occurring while changing the various parameters such as the mean time delay, noise, and repulsion radius were studied.

\begin{figure}[htbp]
\includegraphics[width=3in]{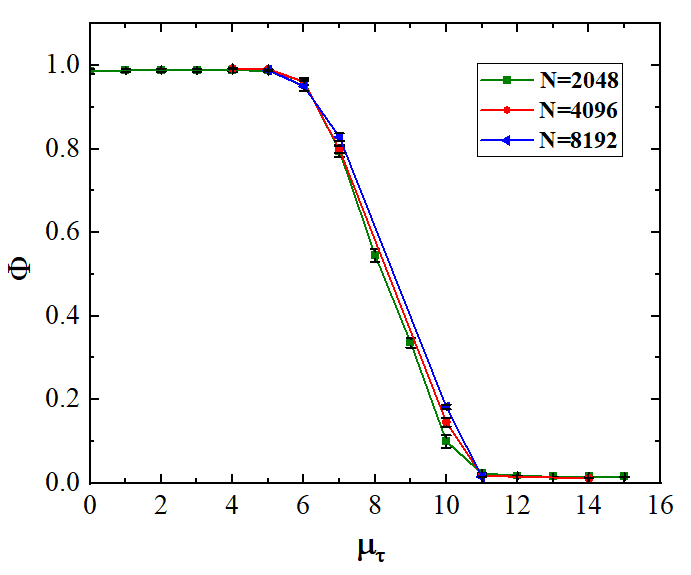}
\caption{A graph of normalized order parameter vs mean time delay for various $N$. $\rho=1.0$, $R =1.0$, $c_{align}=c_{rep}=1.0$, $r^{rep}=0.42$, $v_o=0.01$, and $\eta=0.31$. The phase transition curve is independent of the system sizes.}
\label{Fig2}
\end{figure}

Before deciding which value for N would be optimal to choose for the majority of runs, we looked at the dependency of the order parameter versus delay curves for a set of parameters and for a variety of N values. The results for 3 different numbers of particles, $N =2048$, $4096$, and $8192$ are displayed in Figure \ref{Fig2}  showing the transition from the ordered to the disordered states as a function of the mean time delay. From this figure we conclude that there is only a very weak dependence of the behaviour on the particle number $N$  for $N>2048$ so - unless otherwise stated - the number of particles in our simulations is 2048. Throughout this paper we report on behaviour in the stationary state of the system which is obtained after the order parameter values saturate (start to fluctuate around the given value)

\begin{figure}[htbp]
\includegraphics[width=3in]{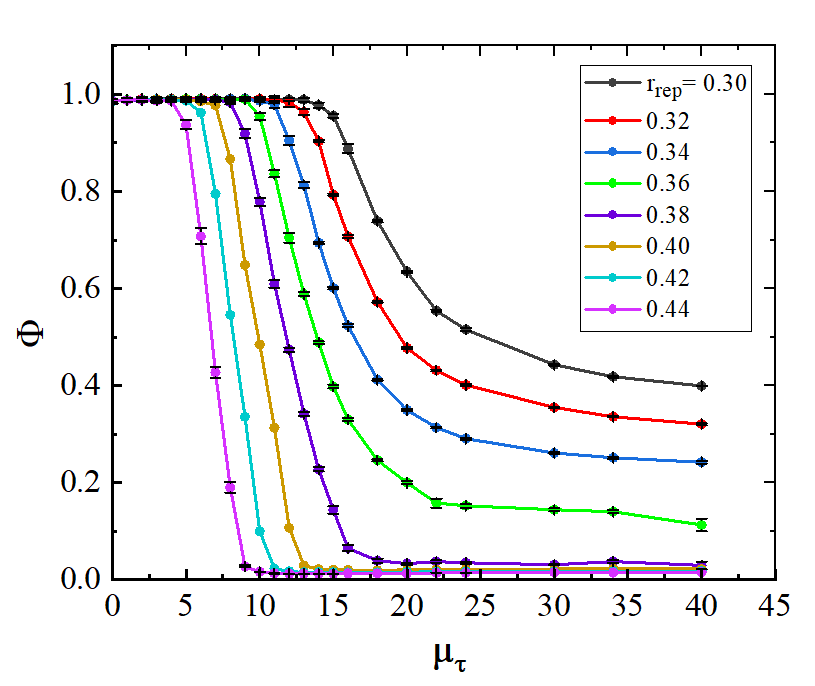}
\caption{Order parameter as a function of mean time delay for various repulsion radii ranging from $0.30$ to $0.44$. The parameters used are: $\rho=1.0$, $R =1.0$, $c_{align}=c_{rep}=1.0$, $v_o=0.01$, and $\eta=0.31$. For $r_{rep} \geq0.38$ there is a phase transition from an ordered phase to a fully disordered phase while for $r_{rep}<0.38$ the system eventually reaches a state which is a mixture of groups of fully disordered agents in the background of orderly moving other particles.}
\label{Fig3}
\end{figure}

To illustrate the effect of delays on the level of coherence of the motion of particles the order parameters as a function of $\mu_{\tau}$ were plotted for various repulsion radii in Figure \ref{Fig3} for moderate level of noise. This figure demonstrates a behaviour that is very specific to our system. In spite of the simplicity of our model it displays an atypical sequence of transitions involving two distinct scenarios and a collective motion phase which has not yet been observed in such simple models. Figure \ref{Fig3} indicates these kinds of transitions: i)  For relatively large repulsion radii (i.e., when two times $r_{rep}$ is close to the alignment radius the system goes through a change from a completely ordered to a fully disordered state as the average delay time is increased. However, ii) for smaller repulsion radii the system - although is also ordered for small delays - does not become fully disordered even if the average delay is substantially increased. In this latter case, the system is in a mixed state of nearly perfectly ordered and fully disordered patches. 

\begin{figure}[htbp]
\includegraphics[width=5.5in]{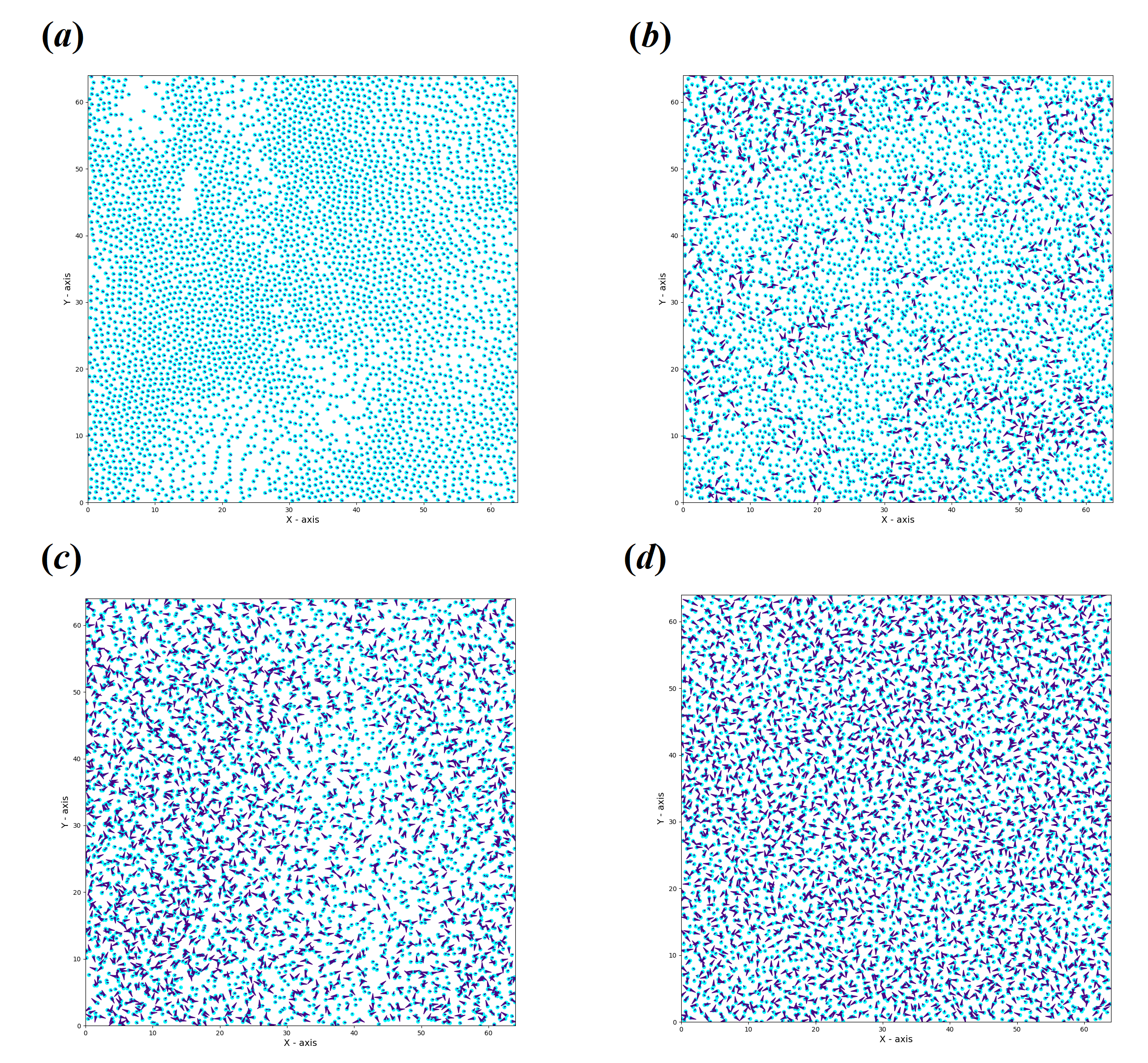}
\caption{Four screenshots of different states of the system for $N=4096$. In all shots $\rho=1.0$, $R =1.0$, $c_{align}=c_{rep}=1.0$, $v_o=0.01$, and $\eta=0.31$.  The instantaneous velocities are indicated by blue vectors of length proportional to their magnitude. $\mathit{(a)}$ The fully ordered state for small delay time, $\mu_{\tau}=5.0$ and $r_{rep}=0.4$ after $5000$ time steps. $\mathit{(b)}$ Mixed state for $\mu_{\tau}=18$ and $r_{rep}=0.3$ after $6000$ time steps. $\mathit{(c)}$ Mixed state for $\mu_{\tau}=35$ and $r_{rep}=0.3$ after $6000$ time steps.$\mathit{(d)}$ The fully disordered state for $\mu_{\tau}=15$ and $r_{rep}=0.4$ after $6000$ time steps.}
\label{Fig4}
\end{figure}

The transition i) to the fully disordered state is accompanied by a sharp decrease in the order parameter indicating a phase change that is similar to a first-order transition, but the decrease is less abrupt and leaves open the possibility of a continuous transition with an exponent being $-1$. The transition to the new, mixed state does not seem to have an analogy with ordinary phase transitions, for example, the value of the order parameter in the large delay limit is dependent on the repulsion radius and thus, does not allow the definition of a single-parameter dependent order parameter displaying values from $1$ to $0$. In Figure \ref{Fig4} four example screenshots of the above-mentioned three states are shown. The videos of the corresponding simulations can be found in Supplementary movies $(2-5)$. Two other videos for $N=16384$ particles with similar parameters i.e. $\rho=1.0$, $R =1.0$, $c_{align}=c_{rep}=1.0$, $v_o=0.01$, and $\eta=0.31$, and for $(r_{rep}=0.38$, $\mu_{\tau}=5)$, and $(r_{rep}=0.32$, $\mu_{\tau}=20)$ is found in Supplementary movies $6$ and $7$ respectively.

\begin{figure}[htbp]
\includegraphics[width=3in]{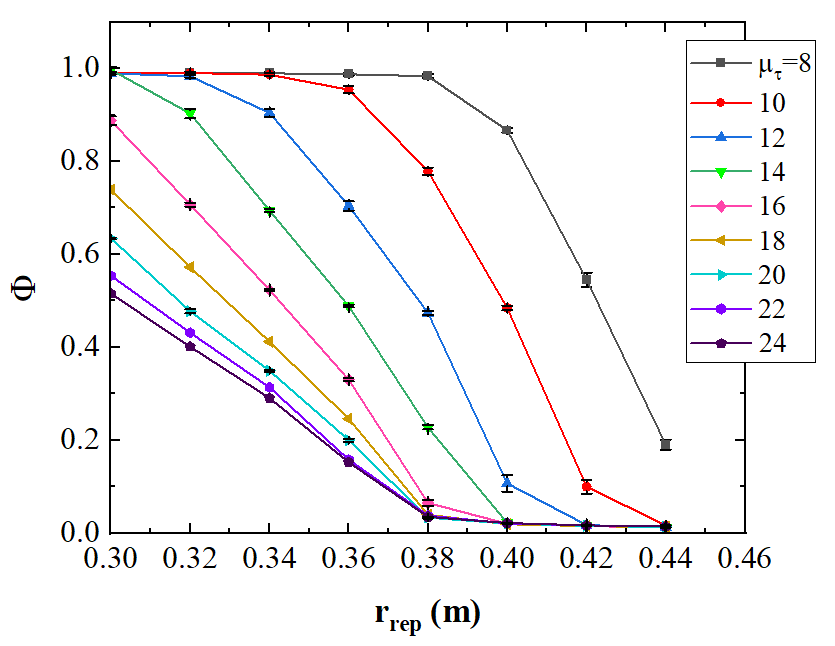}
\caption{Order parameter as a function of $r_{rep}$ for various mean delay time  $\mu_{\tau}$. The parameters used are: $\rho=1.0$, $R =1.0$, $c_{align}=c_{rep}=1.0$, $v_o=0.01$, and $\eta=0.31$. Note that for large delays and relatively small repulsion radii the system stays disordered while it displays ordered collective motion for short average delay times.}
\label{Fig5}
\end{figure}

The two main control parameters we consider in this work are the average delay and the repulsion radius. (Although the model has several more parameters, see Table \ref{table:1}, we mostly used for the other parameters either a simple trivial value, e.g., 1, or some intermediate value. e.g., for the noise, which is far from the given parameter resulting in complex behaviour.) In Figure \ref{Fig3} we showed the behaviour of the order parameter as a function of delay, next we also looked at how the order parameter changes if for a set of fixed delays the radius of repulsion $r_{rep}$ is increased. Figure \ref{Fig5} demonstrates that, as expected, increasing the repulsion radius results in a completely disordered state when $r_{rep}$ approaches 0.5 being half of the alignment radius (2 times $r_{rep}$ getting close to $R=1$). Increased disorder follows from two effects: first, the frequency of "collisions" becomes larger as the effective density grows, and second, the area within which it is only the alignment term which acts (within the annulus between the two circles of radii $r_{rep}$ and $R=1$).

\begin{figure}[htbp]
\includegraphics[width=3in]{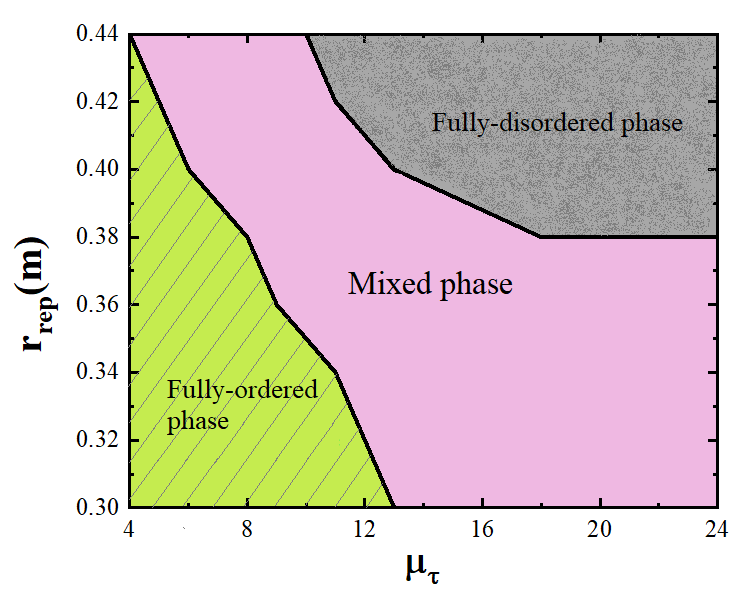}
\caption{The phase diagram for fully ordered, fully disordered, and mixed states with respect to the mean time delay $\mu_{\tau}$ and the repulsion radius $r_{rep}$ for the two-dimensional model of flocking with delay and repulsion.  The other parameters used in the related simulations were; noise: $\eta=0.31$, density: $\rho=1.0$, alignment radius $R=1.0$, repulsion coefficient: $c_{rep}=c_{align}=1.0$, and default velocity $v_o=0.01$.} 
\label{Fig6}
\end{figure}
An important motivation for obtaining the curves in Figure \ref{Fig5} is that together with those displayed in Figure \ref{Fig3} we can construct an approximate phase diagram shown in Figure \ref{Fig6} depicting the regions of the two control parameters $\mu_{\tau}$ and  $r_{rep}$ for which a given collective motion state (Fully ordered, Mixed and Fully-disordered) takes place. What Figure \ref{Fig6} demonstrates is that the regions of the two main control parameters for which all three phases can be observed are quite limited and the extension of the region corresponding to the mixed phase dominates the behaviour.

\section{Conclusions}

Although macroscopic self-propelling (active) units of a system are typically subject to interactions consisting of both a repulsion term and the effect of time delays (reaction times) our simulation is the first study systematically addressing the question of how these two factors affect the collective motion of the units. In the limit of small delays and small repulsion, we find, as expected, fully ordered, coherent motion in the system. Similarly, although less trivially, we obtain a completely disordered motion state for large delays and large repulsion radii. However, in addition to these, we have also found a new kind of collective motion pattern: disordered patches embedded into a globally ordered flow.

We argue that this complex behaviour can be understood by considering competing effects on the behavior: As it is also illustrated by our \ref{Fig1} (a), collisions in the presence of delays can result in trajectories of the particles diverging by an angle being larger than the one along which they approached each other. This mechanism results in an increased disorder and, through a local increase of the magnitude of the "scattered" particles results in an increased pushing of the neighbours deeper into their repulsion area which, in turn, results in further sharp turns of the trajectories. This is an instability which would result in completely disorderly motion without any stabilizing effect. However, delay times - somewhat paradoxically, but in a formerly shown manner (see \cite{Piwowarczyk2019, Holubec2021}) - may result in making the collective motion more coherent. Perhaps the most straightforward interpretation of this phenomenon is to look at very large delays as a limit in which the system behaves quite like its mean-field version without delay (only already distant particles interact). The interplay of the instability due to abrupt collisions and the stabilizing effect of long delays results in a mixed phase containing parts in which either the instability (resulting in disorder)  or the stabilizing long delay dominate the local behaviour. It should be noted that an intermediate value of the order parameter in the present case is not analogous to a partially ordered system of purely aligning particles because in the latter case the level of disorder - due to noise - is distributed homogeneously through the system. 

\section{Acknowledgements}

The authors are grateful to Fahimeh Najafi and M\'{a}t\'{e} Nagy for helpful discussions. This research was partially supported by the following grants: (Hungarian) National Research, Development and Innovation Office grants No: K128780 and SNN 139598. It has also received funding from the European Union’s Horizon 2020 research and innovation programme under the Marie Skłodowska-Curie grant agreement No 955576.

 \bibliographystyle{elsarticle-num} 
 \bibliography{cas-refs}

\begin{thebibliography}{10}
\expandafter\ifx\csname url\endcsname\relax
  \def\url#1{\texttt{#1}}\fi
\expandafter\ifx\csname urlprefix\endcsname\relax\def\urlprefix{URL }\fi
\expandafter\ifx\csname href\endcsname\relax
  \def\href#1#2{#2} \def\path#1{#1}\fi

\bibitem{Vicsek2012}
T.~Vicsek, A.~Zafeiris, Collective motion, Phys. Rep. 517 (2012) 71--140.

\bibitem{Marchetti2013}
M.~Marchetti, J.~Joanny, S.~Ramaswamy, T.~Liverpool, J.~Prost, M.~Rao,
  R.~Simha, Hydrodynamics of soft active matter, Rev. Mod. Phys. 85 (2013)
  1143.

\bibitem{Zhang2010}
H.~P. Zhang, A.~Be'er, E.~L. Florin, H.~L. Swinney, Collective motion and
  density fluctuations in bacterial colonies, Proc. Natl. Acad. Sci. U. S. A.
  107 (2010) 13626--13630.

\bibitem{Mehes2014}
E.~M\'{e}hes, T.~Vicsek, Collective motion of cells: from experiments to
  models, Integr. Biol. 6 (2014) 831--854.

\bibitem{Buhl2006}
J.~Buhl, I.~Sumpter, D. J.and~Couzin, J.~Hale, E.~Despland, E.~R. Miller,
  S.~Simpson, From disorder to order in marching locusts, Science 312 (2006)
  1402--1406.

\bibitem{Lopez2012}
U.~Lopez, J.~Gautrais, I.~D. Couzin, G.~Theraulaz, From behaviuoral analyses to
  models of collective motion in fish schools, Interface Focus 2 (2012)
  693--707.

\bibitem{Bajec2009}
I.~Bajec, F.~Heppner, Organized flight in birds, Animal Behaviour 78 (2009)
  777--789.

\bibitem{Moussaid2010}
M.~Moussaid, N.~Perozo, S.~Garnier, D.~Helbing, G.~Theraulaz, The walking
  behavior of pedestrian social groups and its impact on crowd dynamics, PloS
  one 5 (2010) e10047.

\bibitem{Doostmohammadi2018}
A.~Doostmohammadi, J.~Ign\'{e}s-Mullol, J.~M. Yeomans, F.~Sagu\'{e}s, Active
  nematics, Nat. Commun. 9 (2018) 3246.

\bibitem{Kudrolli2010}
A.~Kudrolli, Concentration dependent diffusion of self-propelled rods, Phys.
  Rev. Lett. 104 (2010) 088001.

\bibitem{Viragh2014}
C.~Vir\'{a}gh, G.~V\'{a}s\'{a}rhelyi, N.~Tarcai, T.~Sz\"{o}r\'{e}nyi,
  G.~Somorjai, T.~Nepusz, T.~Vicsek, Flocking algorithm for autonomous flying
  robots, Bioinspiration Biomimetics 9 (2014) 025012.

\bibitem{Deutsch2012}
A.~Deutsch, G.~Theraulaz, T.~Vicsek, Collective motion in biological systems,
  Interface Focus 2 (2012) 689--692.

\bibitem{Westley2018}
P.~A. Westley, A.~M. Berdahl, C.~J. Torney, D.~Biro, Collective movement in
  ecology: from emerging technologies to conservation and management, Philos.
  Trans. R. Soc. B 373 (2018) 20170004.

\bibitem{Sumpter2010}
D.~J.~T. Sumpter, Collective animal behavior, Princeton University Press
  (2010).

\bibitem{Cont2000}
R.~Cont, J.~Bouchaud, Herd behavior and aggregate fluctuations in financial
  markets, Macroecon. Dyn. 4 (2000) 170--196.

\bibitem{Helbing2000}
D.~Helbing, I.~Farkas, T.~Vicsek, Simulating dynamical features of escape
  panic, Nature 407 (2000) 487--490.

\bibitem{Couzin2009}
I.~D. Couzin, Collective cognition in animal groups, Trends in Cognitive
  Sciences 13 (2009) 36--43.

\bibitem{Vasarhelyi2018}
G.~V\'{a}s\'{a}rhelyi, C.~Vir\'{a}gh, G.~Somorjai, T.~Nepusz, A.~E. Eiben,
  T.~\\~Vicsek, Optimized flocking of autonomous drones in confined
  environments, Sci. Robot. 3 (2018) 3536.

\bibitem{Brambilla2013}
M.~Brambilla, E.~Ferrante, M.~Birattari, M.~Dorigo, Swarm robotics: A review
  from the swarm engineering perspective, Swarm Intell. 7 (2013) 1--41.

\bibitem{Turgut2008}
A.~Turgut, H.~Çelikkanat, F.~G\"{o}kçe, E.~Şahin, Self-organized flocking in
  mobile robot swarms, Swarm Intell. 2 (2008) 97--120.

\bibitem{Aoki1982}
I.~Aoki, A simulation study on the schooling mechanism in fish, Bull. Japan.
  Soc. Sci. Fish 48 (1982) 1081--1088.

\bibitem{Gregoire2003}
G.~Gr\'egoire, H.~Chat\'e, Y.~Tu, Moving and staying together without a leader,
  Phys. D: Nonlinear Phenom. 181 (2003) 157--170.

\bibitem{Nagy2010}
M.~Nagy, Z.~\'{A}kos, D.~Biro, T.~Vicsek, Hierarchical group dynamics in pigeon
  flocks, Nature 464 (2010) 890--893.

\bibitem{Couzin2005}
I.~D. Couzin, J.~Krause, N.~R. Franks, S.~A. Levin, Effective leadership and
  decision-making in animal groups on the move, Nature 433 (2005) 513--516.

\bibitem{Chate2006}
H.~Chat\'e, F.~Ginelli, R.~Montagne, Simple model for active nematics:
  Quasi-long-range order and giant fluctuations, Phys. Rev. Lett. 96 (2006)
  180602.

\bibitem{Narayan2007}
V.~Narayan, S.~Ramaswamy, N.~Menon, Long-lived giant number fluctuations in a
  swarming granular nematic, Science 317 (2007) 105--108.

\bibitem{Cisneros2007}
H.~Cisneros, R.~Cortez, C.~Dombrowski, R.~E. Goldstein, J.~O. Kessler, Fluid
  dynamics of self-propelled microorganisms, from individuals to concentrated
  populations, Exp. Fluids 43 (2007) 737--753.

\bibitem{Sokolov2009}
A.~Sokolov, R.~Goldstein, F.~Feldchtein, I.~Aranson, Enhanced mixing and
  spatial instability in concentrated bacterial suspensions, Phys. Rev. E 80
  (2009) 031903.

\bibitem{Wolgemuth2008}
C.~W. Wolgemuth, Collective swimming and the dynamics of bacterial turbulence,
  Biophysical Journal 95 (2008) 1564--1574.

\bibitem{Kudrolli2008}
A.~Kudrolli, G.~Lumay, D.~Volfson, L.~S. Tsimring, Swarming and swirling in
  self-propelled polar granular rods, Phys. Rev. Lett. 100 (2008) 058001.

\bibitem{Ihle2011}
T.~Ihle, Kinetic theory of flocking: Derivation of hydrodynamic equations,
  Phys. Rev. E 83 (2011) 030901.

\bibitem{Mishra2010}
S.~Mishra, A.~Baskaran, M.~C. Marchetti, Fluctuations and pattern formation in
  self-propelled particles, Phys. Rev. E 81 (2010) 061916.

\bibitem{Tarcai2011}
N.~Tarcai, C.~Vir\'agh, D.~\'Abel, M.~Nagy, P.~V\'arkonyi, G.~V\'as\'arhelyi,
  T.~Vicsek, Patterns, transitions and the role of leaders in the collective
  dynamics of a simple robotic flock, J. Stat. Mech.: Theory Exp. 4 (2011)
  P04010.

\bibitem{Helbing2009}
D.~Helbing, Traffic and related self-driven many-particle systems, Rev. Mod.
  Phys. 73 (2009) 1067.

\bibitem{Reynolds1987}
C.~Reynolds, Flocks, herds, and schools: A distributed behavioral model, In
  Computer Graphics 21 (1987) 25--34.

\bibitem{Vicsek1995}
T.~Vicsek, A.~Czirok, E.~Ben-Jacob, I.~Cohen, O.~Shochet, Novel type of phase
  transition in a system of self-driven particles, Phys. Rev. Lett. 75 (1955)
  1226.

\bibitem{Toner1995}
J.~Toner, Y.~Tu, Long range order in a two-dimensional dynamical xy model,
  Phys. Rev. Lett. 75 (1995) 4326.

\bibitem{Shimoyama1996}
N.~Shimoyama, K.~Sugawara, T.~Mizuguchi, Y.~Hayakawa, M.~Sano, Collective
  motion in a system of motile elements, Phys. Rev. Lett. 76 (1996) 3870.

\bibitem{Couzin2002}
I.~Couzin, J.~Krause, R.~James, G.~Ruxton, N.~Franks, Collective memory and
  spatial sorting in animal groups, J. Theor. Biol. 218 (2002) 1--11.

\bibitem{Toner2005}
J.~Toner, Y.~Tu, S.~Ramaswamy, Hydrodynamics and phases of flocks, Ann. Phys.
  318 (2005) 170--244.

\bibitem{Cucker2007}
F.~Cucker, S.~Smale, Flocks, herds, and schools: A distributed behavioral
  model, IEEE Trans. Automat. 52 (2007) 852--862.

\bibitem{Derzsi2009}
A.~Derzsi, G.~Sz\"{o}l\"{o}si, T.~Vicsek, Most minimal spp model, URL
  http://hal. elte. hu/~ vicsek/SPP-minimal (2009).

\bibitem{Cavagna2018}
A.~Cavagna, I.~Giardina, T.~Grigera, The physics of flocking: Correlation as a
  compass from experiments to theory, Phys. Rep. 728 (2018) 1--62.

\bibitem{Chandler1958}
R.~E. Chandler, R.~Herman, E.~W. Montroll, Traffic dynamics: studies in car
  following, Oper. Res. 6 (1958) 165--184.

\bibitem{Kometani1958}
E.~I. J.~I. Kometani, T.~S. U. N.~A. Sasaki, On the stability of traffic flow
  (report-i), J. Oper. Res. Soc. Jpn. 2 (1958) 11--26.

\bibitem{Herman1959}
R.~Herman, E.~W. Montroll, R.~B. Potts, R.~W. Rothery, Traffic dynamics:
  analysis of stability in car following, Oper. Res. 7 (1959) 86--106.

\bibitem{Macnab1972}
R.~M. Macnab, D.~E. Koshland~Jr, The gradient-sensing mechanism in bacterial
  chemotaxis, Proc. Natl. Acad. Sci. U.S.A. 69 (1972) 2509--2512.

\bibitem{Mackey1977}
M.~C. Mackey, L.~Glass, Oscillation and chaos in physiological control systems,
  Science 197 (1977) 287--289.

\bibitem{Ikeda1979}
K.~Ikeda, Multiple-valued stationary state and its instability of the
  transmitted light by a ring cavity system, Opt. commun. 30 (1979) 257--261.

\bibitem{Landsman2007}
A.~S. Landsman, I.~B. Schwartz, Synchronized dynamics of cortical neurons with
  time-delay feedback, Nonlinear Biomed. Phys. 1 (2007) 1--9.

\bibitem{Carr2006}
T.~W. Carr, I.~B. Schwartz, M.~Y. Kim, R.~Roy, Delayed-mutual coupling dynamics
  of lasers: scaling laws and resonances, SIAM J. Appl. Math. 5 (2006)
  699--725.

\bibitem{Forgoston2008}
E.~Forgoston, I.~B. Schwartz, Delay-induced instabilities in self-propelling
  swarms, Phys. Rev. E 77 (2008) 035203.

\bibitem{Sun2014}
Y.~Sun, W.~Lin, R.~Erban, Time delay can facilitate coherence in self-driven
  interacting-particle systems, Phys. Rev. E 90 (2014) 062708.

\bibitem{Mijalkov2016}
M.~Mijalkov, A.~McDaniel, J.~Wehr, G.~Volpe, Engineering sensorial delay to
  control phototaxis and emergent collective behaviors, Phys. Rev. X 6 (2016)
  011008.

\bibitem{Casas2018}
V.~Casas, A.~Mitschele-Thiel, On the impact of communication delays on uavs
  flocking behavior, IEEE Wireless Communications and Networking Conference
  Workshops (WCNCW) (2018) 67--72.

\bibitem{Piwowarczyk2019}
R.~Piwowarczyk, T.~Selin, M.and~Ihle, G.~Volpe, Influence of sensorial delay on
  clustering and swarming, Phys. Rev. E 100 (2019) 012607.

\bibitem{Holubec2021}
V.~Holubec, D.~Geiss, S.~Loos, K.~Kroy, F.~Cichos, Finite-size scaling at the
  edge of disorder in a time-delay {V}icsek model, Phys. Rev. Lett. 127 (2021)
  258001.

\bibitem{Hemelrijk2015}
C.~Hemelrijk, L.~van Zuidam, H.~Hildenbrandt, What underlies waves of agitation
  in starling flocks, Behav. Ecol. Sociobiol. 69 (2015) 755--764.

\bibitem{Islam2023}
M.~Islam, I.~Faruque, Insect visuomotor delay adjustments in group flight
  support swarm cohesion, Sci. Rep. 13 (2023) 6407.

\end{thebibliography}





\end{document}